\documentclass[12pt]{article}
\usepackage{amsmath,amssymb}

\usepackage[dvips]{graphicx}
\DeclareGraphicsExtensions{.eps}

\newtheorem{theorem}{Theorem}

\title{Unstable even-parity eigenmodes of the regular static SU(2) Yang-Mills-dilaton solutions}

\author{O.I. Streltsova\thanks{E-mail: strel@jinr.ru}, E.E. Donets, E.A. Hayryan, \\ D.A. Georgieva, and T.L. Boyadjiev}

\date{}

\begin{document}

\maketitle

\vspace{-1.cm}

\begin{center}
{\it Joint Institute for Nuclear Research, 141980 Dubna, Russia}
\end{center}

\begin{abstract}

In this paper we obtain unstable even-parity eigenmodes to the
static regular spherically symmetric solutions of the SU(2)
Yang-Mills-dilaton coupled system of equations in $3+1$ Minkowski
space-time. The corresponding matrix Sturm-Liouville problem is
solved numerically by means of the continuous analogue of Newton's
method. The method, being the powerful tool for solving both
boundary-value and Sturm-Liouville problems, is described in
details.

\end{abstract}

\section{Introduction}

Regular stationary/static solutions of nonlinear field equations
play a very important role for understanding of non-perturbative
aspects of field dynamics in various modern field models,
including gravity. During the last decade since Bartnik and
McKinnon's discovery \cite{bk} of the regular particle-like
solutions to the coupled system of the Einstein-Yang-Mills (EYM)
equations there was a growing interest in revealing their possible
physical and mathematical significance \cite{gv} -- \cite{GS}. At
the same time, the qualitatively similar (essentially non-Abelian
ones) regular solutions were found in other various field models,
i.e. in the coupled system of the Yang-Mills-dilaton (YMD)
equations \cite{LM}, in the coupled system of the
Einstein-Yang-Mills-dilaton (EYMD) equations \cite{dg1}--
\cite{tm}, in more sophisticated stringy inspired field models
\cite{dg3} etc (see Ref.\ \cite{rep} for a review).

The main physical and mathematical properties of the mentioned
solutions are determined by the nonlinear nature of the SU(2)
Yang-Mills (YM) field which is a substantial part of all these
models. All the mentioned coupled systems of equations admit
spherically symmetric particle-like solutions in which the
Yang-Mills field configuration is bounded by gravity (EYM), or by
the scalar dilaton field (YMD), or by both of them (EYMD) in
unstable equilibrium. (Note, pure YM field in the $3+1$ Minkowski
space-time is self-repulsive and no regular pure YM configurations
exist.) In terms of the linear perturbation analysis these
solutions have even-parity unstable modes \cite{SZ}, \cite{LM},
\cite{LM2}, which are called gravitating (EYM) or dilatonic (YMD)
unstable modes.

On the other hand, in the case of a spherical symmetry the SU(2)
Yang-Mills self-interaction potential has a double-well shaped
form and its two distinct minima correspond to the topologically
distinct YM vacua. The relevant YM magnetic function must tend to
the vacuum values at the origin and at the spatial infinity for
all regular solutions. It was shown some those regular solutions
could play a role at ultramicroscopic distance analogous to that
of electroweak sphalerons \cite{gv} due to existence of the
odd-parity Yang-Mills negative modes \cite{vg} (which are called
sphaleronic unstable modes) and the corresponding fermion zero
modes, which are responsible for the anomalous fermion production
\cite{GS}.

However, in this paper we concentrate on study of the even-parity
unstable modes in the YMD system, since their significance is
inspired by an interesting story, related to the singularity
formation in nonlinear evolution equations. Indeed, it has been
discovered in the numerical studies of the massless fields
collapse \cite{ch1}, there are two types of the collapse behavior.
Type-II behavior is characterized by the mass gap absence in the
black hole spectrum hence the black holes with an arbitrary small
mass can be obtained in this way. Later on, the Type-I behavior
was observed as well \cite{ch2}, \cite{sky}. It takes place if the
considered system of Einstein-matter equations admits regular
static finite energy asymptotically flat solutions. In this case
the smallest black hole has a finite mass which is equal to the
mass of the lowest static solution. In the EYM system the lowest
static solution is the $N=1$ Bartnick-McKinnon (BK) solution and
its mass determines the smallest mass of the formed black hole in
this type of a collapse. $N=1$ BK solution is occurred to be an
intermediate attractor, which the collapsing solution should
attain in order to turn to the Type-I black hole formation
scenario \cite{ch2}, \cite{gund}, \cite{hirs}. Moreover, $N=1$ BK
solution is occurred to be a threshold configuration, which
separates the black hole formation and dispersion scenarios. The
revealing of the threshold nature remains an unresolved problem
and the studies of the static configurations decay via their
proper unstable eigenmodes seem to be a very natural step towards
understanding the threshold nature.

Thus, the goal of our present paper is to obtain unstable
eigenmodes of the regular static solutions in the YMD system. It
has been realized recently that the singularity (black hole)
formation in gravity and blow-up in nonlinear wave equations share
many common features \cite{b1}, \cite{b2}, \cite{b3}, \cite{lin},
since one deals with the same class of the supercritical evolution
PDE's \cite{b22}. It was shown in Ref. \cite{PRD}, a coupled
system of Yang-Mills-dilaton equations in $3+1$ Minkowski space is
also a supercritical system in PDE terminology and it does exhibit
blow-up (singularity formation) behavior, which is, in principle,
equivalent to the black hole formation. So, we study the unstable
modes of the static solutions in the YMD system, which is
representative for a wide class of a supercritical evolution
PDE's, including  self-gravitating systems.

Our main tool to attack the problem numerically is a continuous
analogue of the Newton's method (CANM), which is a really powerful
tool for solving both boundary-value and Sturm-Liouville problems.
The key point of the method is an introduction of formal evolution
parameter $t$ and a corresponding reformulation of the problem in
terms of a new system of PDE evolution equations. If necessary
conditions are satisfied, the solution of the evolution problem
approaches the solution of the initial (boundary-value or
Sturm-Liouville) ODE problem as $t \rightarrow +\infty$.

The paper is organized as follows. In the next Section we briefly
remind the main features of the regular static spherically
symmetric solutions to the coupled system of the SU(2)
Yang-Mills-Dilaton equations. In the third Section the iterative
scheme, based on the continuous analogue of Newton's method (CANM)
is introduced. The CANM method is used for numerical solution of
the matrix Sturm-Liouville problem, which describes (unstable)
eigenmodes of the static regular YMD solutions. The fourth Section
contains the results of the numerical experiments
--- the eigenvalues and the eigenfunctions of the mentioned
Sturm-Liouville problem. We conclude with some remarks. A more
detailed description of the CANM method is presented in the
Appendix.

\section{Basic equations.}

\hspace*{\parindent} Static spherically symmetric regular solutions
of a coupled system of Yang-Mills-dilaton (YMd) equations
in $3+1$ Minkowki space-time were obtained and analyzed in
Ref. \cite{LM}.
In this Section we briefly remind their results.

A coupled system of YMd fields is given by the action:
\begin{equation} \label{action}
    S=\frac{1}{4\pi}\int \left(\frac{1}{2}\,(\partial\Phi)^2 -
    \frac{\exp\{k\Phi\}}{4g^2} F^{a\mu\nu} F^{a}_{\mu\nu}\right) \,d^3x\,dt,
\end{equation}
\noindent where $\Phi$ is the dilaton field, $F^{a\mu\nu}$ -- the
Yang-Mills field, $k$ and $g$  are dilaton and gauge coupling
constants, respectively. Purely magnetic ansatz for the static
$SU(2)$ spherically symmetric YM potential has a form:
\begin{equation}
     A^a_t = 0 \quad \quad A^a_i= \epsilon_{aik}\frac{x^k}{r^2}\,
     \left[f(r)-1 \right], \label{2}
\end{equation}
\noindent  hence the desired static solutions are described in
terms of two independent functions: $f(r)$~ -- the YM function and
$\Phi(r)$~-- the dilaton  function. After substitution \eqref{2}
into \eqref{action} and rescaling $\Phi\to\Phi/k,\, r\to (k/g)r,\,
t\to (k/g)t$, and $S \to g*k S$, the dependence on two parameters
$k$ and $g$ is effectively vanished. The integration over the
angular variables gives the reduced action:
\begin{equation} \label{actionI}
    S = -\int\limits_0^\infty \left\{\frac{1}{2}\,r^2 \Phi\,'\,^2 +
    \exp\{\Phi\}\left[f\,'\,^2 + \frac{(f^2-1)^2} {2r^2} \right]\right\}\,dr\,,
\end{equation}
\noindent where prime stands for  derivatives with respect  to the
radial variable $r$. The corresponding field equations
\begin{subequations}\label{stat}
    \begin{eqnarray}
        && f\,''+f\,'\Phi\,' = -\frac{f(1-f^2)}{r^2}\,,  \label{5a} \\
        &&\Phi\,''+\frac{2\Phi\,'}{r}=\frac{e^\Phi}{r^2} \left[ f\,'^2
+ \frac{(f^2-1)^2} {2r^2} \right]\,, \label{5b}
    \end{eqnarray}
\end{subequations}
are obtained by a variation of the action (\ref{actionI}) in respect to
the functions $f(r)$ and $\Phi(r)$ under the following boundary conditions:
\begin{subequations} \label{bcstat}
    \begin{gather}
        f(0) = \pm 1, \quad f(\infty) = \pm 1\,, \\
        \Phi\,'(0) = 0\,, \quad \Phi\,'(\infty) = 0\,.
    \end{gather}
\end{subequations}
\noindent The invariance of the equations \eqref{stat} in respect
to the  replacement $f\rightarrow - f $ allows one to put $f(0)=1$
without loss of generality. The transformation $\Phi \rightarrow
\Phi+\lambda$, $r\rightarrow r\exp\{-\lambda/2\}$ with
$\lambda=\mbox{const}$ does not change the equations either, hence
one can put $\Phi_0=0$.

The system of nonlinear equations \eqref{stat} has two singular
points $r=0$ and $r=\infty$ and the solutions are supposed to be
bounded at these points.

Regular solutions at the point $r=0$ admit the following simple series expansions:
\begin{subequations}\label{expans0}
    \begin{gather}
        f(r)_{r \to 0} = 1 - b\, r^2+O(r^4)\,, \\
        \Phi(r)_{r\to 0} = \Phi_0 + b^2 r^2 + O(r^4)\,.
    \end{gather}
\end{subequations}
\noindent Here $b$ is a free parameter, that finally provides the boundary
conditions at the origin as:
\begin{subequations}
    \begin{gather}\label{boundary1}
        f(0)=1,\quad  f\,'(0)=0\,,\\
        \Phi(0)=0,\quad \Phi\,'(0)=0.
    \end{gather}
\end{subequations}

The asymptotical behavior of regular
solutions at $r \to \infty$ has the following form:
\begin{eqnarray} \label{expansInf}
    &&f(r)_{r \to \infty}=(-1)^N \left(1-\frac{c}{r}+
\frac{3c^2-c d}{4 r^2}-\frac{11c^3-6 c^2d +c d^2}{20\,r^3}+
O(r^{-4})\right),\nonumber\\ [0.1cm]
    &&\Phi(r)_{r \to \infty}=\Phi_{\infty}-\frac{d}{r}+O(r^{-4}),
\end{eqnarray}
where $c$, $d$, and $\Phi_{\infty}$ are parameters. As a result,
the corresponding boundary conditions at the infinity are
\begin{eqnarray} \label{boundary2}
    \nonumber &&\lim_{r\to \infty}f(r)=(-1)^N ,\quad
    \lim_{r\to \infty}f\,'(r)=0,\\[0.1cm]
    &&\lim_{r\to \infty}\Phi(r)=\Phi_{\infty}, \quad
    \lim_{r\to \infty}\Phi\,'(r)=0,
\end{eqnarray}
\noindent for $N = 1, 2, \ldots$

\begin{figure}
    \centering
    \includegraphics[width=14.cm]{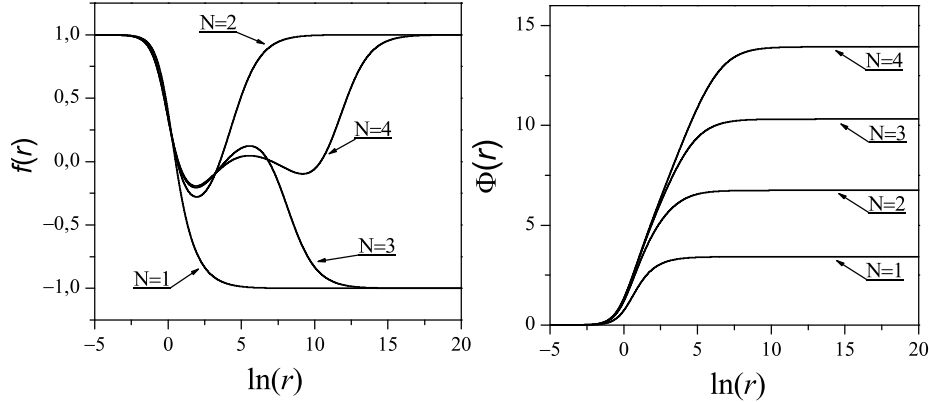}\caption { \footnotesize
Static solutions with $N=1,2,3,4$: YM function $f(r)$ --
    on the left, dilaton function $\Phi(r)$ -- on the right.}
\label{fig:1}
\end{figure}

It is well-known fact that the system of equations \eqref{stat}
has an infinite set of regular solutions, labeled by the number
$N$ of the nodes of YM function $f(r)$ \cite{LM},\cite{Maison}. In
the Ref.\cite{LM} the solutions have been found numerically by use
of the shooting strategy in respect to the free parameter $b$.
Indeed, in order to get solutions it is possible to transform the
boundary value problem \eqref{stat}, \eqref{boundary1},
\eqref{boundary2} into the initial value problem with the initial
conditions \eqref{expans0} so that the desired solutions meet the
asymptotics \eqref{boundary2} at infinity. For the selection of
the parameter $b$ the following key property \cite{lokReg} of the
YM function was used: YM function $f(r)$ is strictly bounded
within the interval $-1\leq f(r)\leq 1$ for all globally regular
solutions of the system \eqref{stat}.

For our present purposes  the solutions of the system \eqref{stat}
have been found using the continuous analogue of Newton's method
\cite{Pyzunin2} (see below Chapter 3). In this approach the
problem is considered as a boundary value problem within some
finite interval $r \in [0,R_\infty ]$, where $0 < R_\infty <
\infty$ is an ``actual infinity''. The boundary conditions can be
written as follows:
\begin{subequations}
    \begin{gather*}
        f(0) = 1, \quad f(R_\infty) = (-1)^N\,,\\
        \Phi\,'(0) = 0\,, \quad \Phi\,'(R_\infty) = 0\,.
    \end{gather*}
\end{subequations}
For numerical realization it is effective to use a boundary
condition of a mixed type
\begin{equation}\label{bcr}
    \frac{1}{R_{\infty}} \left[ f(R_\infty)-
(-1)^N \right] + f\,'(R_\infty) = 0\,,
\end{equation}
which directly follows from the asymptotics \eqref{expansInf}.

The static solutions corresponding to $N=1,2,3,4$ are shown in Fig.\ \ref{fig:1}.

Following the lines of Ref.\ \cite{LM} we shall consider  small
spherically symmetric perturbations of the obtained regular static
solutions $f_N(r),\Phi_N(r)$ of the following form:
\begin{subequations} \label{voz}
    \begin{gather}
        f(r)=f_N(r)+\epsilon \,\exp\{-\Phi_N/2\}\, v(r)\, \exp\{i\omega t\}\,,\\
        \Phi(r) = \Phi_N(r) + \epsilon\,  \frac{\sqrt{2}}{r}\,u(r)\, \exp\{i\omega t\}.
    \end{gather}
\end{subequations}
Here $\epsilon$ is a small parameter.

After introducing the column of the perturbation functions
$\chi \equiv \left( u, v\right)^T$ (upper index $T$
everywhere sign the transposition) the effective action for
the perturbations
$u(r), v(r)$ in the leading non-vanishing order (~$\epsilon^2$)
has the form
\begin{equation} \label{Eaction}
    {\widetilde S} = \int\limits_0^\infty \left( (\chi^{+})'\, (\chi)' +
p(r) (\chi^{+})' (i\sigma_2) \chi + \chi^{+}\,V(r) \chi -
\omega^2 \chi^{+} \chi \right) \, dr + \omega^2\,,
\end{equation}
where $\sigma_2$ is the corresponding Pauli matrix. Note,  we have
added the term $\omega^2$ into the effective action
\eqref{Eaction} in order to get the normalization condition (see
below) by a formal variation in respect to the
$\lambda=-\omega^2$. The elements of the matrix $V(r)$ and the
function $p(r)$ are expressed in terms of the background static
solution $f_N(r), \Phi_N(r)$ as follows:
\begin{eqnarray}
    \nonumber
    V_{11}(r)&=&\frac{\exp\{\Phi_N\}}{r^2}\left[{f\,'_N}^2+ \frac{(f_N^2-1)^2}{2 r^2}\right], \\
\nonumber
    V_{12}(r)&=&V_{21}(r)= \frac{1}{\sqrt{2}\,r^2} \left[\,r \exp\{\Phi_N/2\}f\,'_N \right]' \,, \\
\nonumber
"    V_{22}(r)&=&\frac{1}{2}\,\Phi\,''_N + \frac{1}{4}\,{\Phi\,'_N}^2 + \frac{3f_N^2-1}{r^2};\\"
    \nonumber
    p(r)&=& - 2\frac{\exp\{\Phi_N/2\}f\,'_N} {\sqrt{2}\,r}.
\end{eqnarray}

 In order to bring the eigenvalue problem to a self-adjoined form,
the extended derivative $D$ is introduced:
\begin{eqnarray}\label{xi}\nonumber
    &&D\chi=\chi\,' -i\,A'\sigma_2\,\chi={u'-A'v\choose v'+A'u}\\[0.1cm]
    &&A(r)=\int\limits_0^{r}\frac{\exp\left[{\phi_N(\xi)/2}\right]
f_N(\xi)_{,\,\xi}} {\sqrt{2}\,\xi}\,d \xi\,.
\end{eqnarray}
Then the effective action (\ref{Eaction}) can be rewritten in the form:
\begin{equation}
    {\widetilde S} = \int\limits_0^\infty \left( D \chi^{+}\,D \chi +
    \chi^{+}\,\tilde{U}(r) \chi - \omega^2 \chi^{+} \chi \right)\, dr +
\omega^2\,.
\end{equation}
The elements of the matrix $\tilde{U}(r)$ are expressed as:
\begin{eqnarray}
    \nonumber
    \tilde{U}_{11}(r)&=&\frac{\exp\{\Phi_N\}}{2\,r^2}\left[{f\,'_N}^2+
    \frac{(f_N^2-1)^2} {r^2} \right], \\
    \tilde{ U}_{12}(r)&=& \tilde{U}_{21}(r) = \frac{1}{\sqrt{2}\,r^2} \,
    \left[ r\, \exp\{\Phi_N/2\} f\,'_N \right]' ,\\
\nonumber
    \tilde{U}_{22}(r)&=&\frac{1}{2}\,\Phi\,''_N + \frac{1}{4}\,{\Phi\,'_N}^2 +
    \frac{3f_N^2-1} {r^2} - \frac{\exp\{\Phi_N\} {f\,'_N}^2} {2\,r^2}\,.
\end{eqnarray}

\begin{figure}
    \includegraphics[width=11.cm]{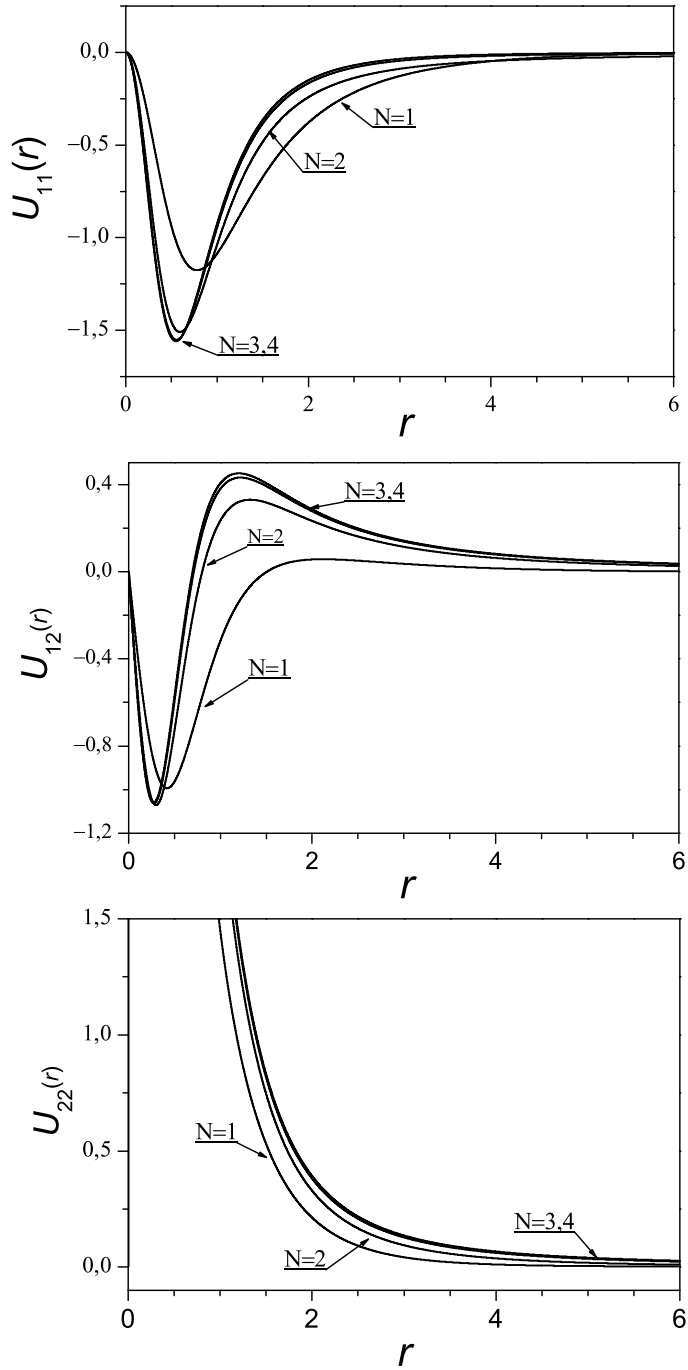}
    \caption{ \footnotesize The matrix elements $U_{ij}$ (\ref{U}):
  $U_{11}(r)$ -- top, $U_{12}(r)$ -- middle, $U_{22}(r)$ -- bottom
which correspond to the static solutions with $N=1,2,3,4$ nodes. }\label{fig:2}
\end{figure}

The gauge transformation
\begin{eqnarray} \label{psidef}
    \chi=\exp\{i\, A(r) \sigma_2\} \,\Psi(r)\,,
\end{eqnarray}
\noindent where $\Psi= \left( \Psi_1, \Psi_2 \right)^T$, allows
one to bring the action to the final self-adjoint form
\begin{equation} \label{sfunc}
    {\widetilde S}=\int\limits_0^\infty \left( (\Psi^{+})'(\Psi)'+
\Psi^{+}\,U (r)\Psi- \omega^2 \Psi^{+} \Psi \right) \, dr + \omega^2\,
\end{equation}
\noindent where the elements of the matrix potential $U(r)$ are
\begin{equation} \label{U}
    U(r)=\exp\{-iA(r)\sigma_2\}\,\tilde{U}(r) \exp\{iA(r)\sigma_2\},
\end{equation}
\noindent The behavior of the matrix elements $U_{ij}$ (\ref{U}),
corresponding to the static solutions with $N=1,2,3,4$ nodes, is
demonstrated on Fig.\ \ref{fig:2}.

Let us put $\omega^2 =-\lambda$, then we get a matrix Sturm-Liouville
problem as a result of the unconditional extremum
conditions of the functional \eqref{sfunc} on the set of the variables
$\left(\Psi_1, \Psi_2, \lambda\right)$:
\begin{equation} \label{ShturLiuv}
    -\Psi\,'' + U(r) \Psi + \lambda\,\Psi = 0\,,
\end{equation}
on the semi-axes  $0 \le r < \infty$ with the boundary conditions:
\begin{equation}\label{GrC}
    \Psi_1\,(0)= 0\,, \quad \Psi\,'_2(0) = 0, \quad \Psi_1\,(\infty)=0\,
, \quad \Psi_2(\infty) = 0.
\end{equation}
and the norm condition
\begin{equation}\label{norma}
    I(\Psi)=\int\limits_0^{\infty}\Psi^{+}\Psi\, d\,r -
    1 \equiv \int\limits_0^{\infty} \left(\Psi_1^2 + \Psi_2^2 \right)\, d\,r - 1=0.
\end{equation}

The linear stability of the static solutions is related to the
spectrum of the eigenvalue problem \eqref{ShturLiuv} --
\eqref{norma}: the negative eigenvalues $\omega^2=-\lambda<0$
correspond to the unstable modes. In the Ref.\ \cite{LM} the
Calogero phase functions method \cite{calogero} has been applied
in order to prove the unstable eigenmodes existence: it was shown
that each static regular solution with $N$ nodes of the YM
function had exactly $N$ unstable eigenmodes in the spectrum of
the Sturm-Liouville problem \eqref{ShturLiuv} -- \eqref{norma}.

The goal of our present paper is to get numerically the unstable
eigenvalues and the corresponding eigenfunctions of the
Sturm-Liouville problem \eqref{ShturLiuv} -- \eqref{norma} with
accuracy, which should be sufficient for further simulation of the
static regular solution decay via their unstable eigenmodes.

\section{Iterative scheme for the Sturm-Liouville problem}

\hspace*{\parindent} In what follows, the elaborated continuous
analogue of Newton's method (CANM) \cite{gavurin} -- \cite{Pyzunin2} was applied to solve our Sturm-Liouville problem.

The first step is to replace the semi-infinite interval  $r \in
[0,\infty)$ by the interval $r\in [0,R_\infty]$ for the numerical
solution of the problem. Here $R_\infty \gg 0$ is the ``actual
infinity''. At the same time the problem \eqref{ShturLiuv} --
\eqref{norma} can be rewritten in the form
\begin{subequations}\label{ShLiuvR}
    \begin{gather}
        E(\Psi,\lambda)= -\Psi\,'' + U \Psi + \lambda\Psi=0,\label{GLR}\\
        G_L(\Psi(0),\Psi'(0),\lambda) = 0\,, \quad G_R(\Psi(R_\infty),
        \Psi'(R_\infty),\lambda) = 0\,,\label{normaR} \\
        I(\Psi)=\int\limits_0^{R_\infty}\Psi^{+} \Psi\, d r - 1 =
        \int\limits_0^{R_\infty}
        \left[ \Psi_1^2(r) + \Psi_2^2(r) \right]\, d r - 1=0,
    \end{gather}
\end{subequations}
where $G_L$ and $G_R$ are boundary conditions (of mixed type, in
general) at the  left ($r=0$) and at the right ($r=R_{\infty}$)
ends of the interval.

The boundary conditions at the left side $G_L$ are already imposed
in \eqref{GrC}, so we just rewrite it as
\begin{eqnarray}\nonumber
    G_{1L}(\Psi(0),\Psi\,'(0),\lambda)&\equiv&\Psi_1(0)=0,\\
    G_{2L}(\Psi(0),\Psi\,'(0),\lambda)&\equiv&\Psi_2'(0)=0. \label{GL0}
\end{eqnarray}
\noindent The consistency of the imposed boundary condition
$G_{1L}=G_{2L}=0$ is confirmed by the local solutions behavior,
obtained from the series expansions of the system \eqref{GLR}
at the vicinity of the origin:
\begin{eqnarray}\label{Uijr0}
    \nonumber
    \mathop {U_{11}(r)}\nolimits_{r \to 0} &=& 2b^3 \left(b - 4\right)r^2 +
    O\left(r^4\right)\,\\ \nonumber
    U_{12}(r)_{r\to 0}&=&  U_{21}(r)_{r\to 0} =
    \frac{4}{3} \,\sqrt{2}\,b^2\left( b-3 \right)r + O(r^{3}),\\\nonumber
    U_{22}(r)_{r\to 0}&=& \frac{2}{r^2}+ 3b\,(b-2)+ \frac{6}{5}\,b^2(4+4b-b^2)\;
    r^2  +O\left(r^{4}\right); \\\nonumber
    \Psi_1(r)_{r\to 0}&=&  \tilde{C}_{1}\,\left(r+\frac{1}{6}\lambda r^3\right)+
    O\left(r^{5}\right)\,,\\
\nonumber
    \Psi_2(r)_{r\to 0}&=& \tilde{C}_2\,r^2+ O\left(r^{4}\right)\,.
\end{eqnarray}
\noindent Here $\tilde{C}_1$ and $\tilde{C}_2$ are free parameters.
The static solutions behavior \eqref{expans0} was used to obtain
the matrix elements $U_{ij}(r)$ asimptotics at $r \rightarrow 0$.

The boundary conditions  $G_R$ at the right end ($r=R_{\infty}$)
can be defined in a similar way, taking into account the local
solutions behavior at $r \rightarrow \infty$. Indeed, using the
asymptotic behavior of the static solutions (\ref{expansInf}) and
the matrix elements $U_{ij}(r)$:
\begin{eqnarray}\label{UijInf}
    \nonumber
    U_{11}(r)_{r\to\infty}&=&\left(\frac{2} {r^2} -
\frac{6c+d}{r^3} \right) \sin^2(A_{\infty}) + O(r^{-4}), \\
    U_{12}(r)_{r\to\infty}&=&  U_{12}(r)_{r\to\infty} =
    \left( -\frac{1}{r^2} + \frac{6c+d} {2\,r^3}\right) \sin(2\,A_{\infty}) +
    O(r^{-4}), \\\nonumber
    U_{22}(r)_{r\to\infty}&=&\left( \frac{2}{r^2}-
    \frac{6c+d}{r^3}\right)\cos^2(A_{\infty}) + O(r^{-4}),
\end{eqnarray}
where the value $A_{\infty}\equiv A(r=\infty)$ (according to \eqref{xi}),
we obtain the local solutions for the functions $\Psi$ given by
the following series expansion:
\begin{eqnarray}\label{PsiInf}
\nonumber
    \Psi_{1}(r)_{r\to\infty}&=&\left[ C_1\left(1 +
    \frac{\sin^2(A_{\infty})} {\sqrt{\lambda}\,r} \right) -
    C_2\frac{\sin(2A_{\infty})}{2\,\sqrt{\lambda}\,r}+
    O(r^{-2})\right]e^{-\sqrt{\lambda}\,r}, \\
    \Psi_{2}(r)_{r\to\infty}&=&\left[ C_2\left(1+ \frac{\cos^2(A_{\infty})}
    {\sqrt{\lambda}\,r} \right) - C_1\frac{\sin(2A_{\infty})}
    {2\,\sqrt{\lambda}\,r} + O(r^{-2}) \right] e^{-\sqrt{\lambda}\,r}.
\end{eqnarray}
\noindent If we restrict our considerations within the terms of
order $O(r^{-1})\,e^{-\sqrt{\lambda}\,r}$, the following couple
$\left(G_{1R}, G_{2R}\right)$ of the boundary conditions at the
right end $r=R_{\infty}$ are given by
\begin{subequations}\label{Gr-2sh}
    \begin{gather}
        G_{1R}(\Psi(R_\infty),\Psi\,'(R_\infty),\lambda) \equiv \Psi_1\,
        '(R_{\infty}) + \sqrt{\lambda}\, \Psi_1(R_{\infty})=0, \\[0.1cm]
        G_{2R}(\Psi(R_\infty),\Psi\,'(R_\infty),\lambda)
        \equiv \Psi_2'(R_{\infty}) + \sqrt{\lambda}\, \Psi_2(R_{\infty}) = 0.
    \end{gather}
\end{subequations}

The problem \eqref{ShLiuvR} is equivalent to the nonlinear
functional equation \cite{Pyzunin1}
\begin{equation}\label{main}
    \varphi (z) \equiv \left( {\begin{array}{*{20}c}
    {E(z)} \hfill  \\
    {G_L (z)} \hfill  \\
    {G_R (z)} \hfill  \\
    {I(z)} \hfill  \\
 \end{array} } \right) = 0,
\end{equation}
where the elements $z \equiv \left(\lambda,\Psi\right)$, $\lambda
\in \mathbb{R}$,  and $\Psi_i(r)$ belong to the set of twice
differentiable functions on the finite interval $[0,R_\infty]$,
which satisfy the boundary conditions \eqref{normaR}.

It is supposed that an isolated solution  $z^* \equiv
\left(\lambda^*,\Psi^* \right)$ of the equation \eqref{main} does
exist. Then, according to the CANM method \cite{gavurin} --
\cite{Pyzunin2}, a formal evolution parameter $t$ is introduced
so, the following evolution equation
\begin{eqnarray}\label{evol}
\frac{d}{dt}\varphi(z(t),t)=-\varphi(z(t),t), \quad 0 < t < \infty\,,
\end{eqnarray}
\noindent
is considered with the initial condition
$$z(0)=z_0\,.$$
It is also supposed that the given initial  data $z_0 \equiv
\left(\lambda_0,\Psi_0\right)$ is chosen in some vicinity of the
exact desired solution $z^*$ in a corresponding functional space.
In Ref. \cite{Pyzunin2} it was shown that if additionally:
\begin{enumerate}
 \item $\varphi(z) $is a smooth function;
 \item the operator $\left[\varphi\,'(z)\right]^{-1}$ exists and is bounded
in some vicinity of the point $z(0)=z_0$;
\end{enumerate}
then the equation \eqref{evol} has a unique local solution $z(t)$
in the vicinity of $z^*$, and
$$\lim_{t\to\infty} z(t)= z^*\,.$$

We assume the above two conditions are fulfilled.

Let us introduce the functions $w(r,t)$ and $\mu(t)$
in the following way:
\begin{subequations}\label{def}
    \begin{gather}\label{det-w}
        w(r,t) = \frac{\partial\,\Psi(r,t)} {\partial t} =
        {w_1(r,t) \choose w_2(r,t)},\\ \label{det-nu}
      \mu(t) = \lambda(t)+\frac{d\,\lambda(t)}{dt}\,.
    \end{gather}
\end{subequations}
Then the equation \eqref{evol} in our particular example is
rewritten in terms  of a couple of equations which allow one to
determine the introduced functions $w(r,t)$ and $\mu(t)$ as
follows:
\begin{subequations}\label{Nv}
    \begin{gather}
        -w\,'' + (U+\lambda I) w = -(-\Psi\,'' + (U +\mu I)\Psi) \,,\\ \label{N-GLR1}
         \quad \frac{d}{d\,t} G_L(\Psi(R_\infty),\Psi'(R_\infty),\lambda)
           =-G_L(\Psi(R),\Psi\,'(R_\infty),\lambda),\\\label{N-GLR}
         \quad \frac{d}{d\,t} G_R(\Psi(R_\infty),\Psi'(R_\infty),\lambda)
           =-G_R(\Psi(R),\Psi\,'(R_\infty),\lambda) \,,\\ \label{N-normaR}
        \int\limits_0^{R_\infty} \left[ \Psi_1(r,t)w_1(r,t)+ \Psi_2(r,t)
        w_2(r,t)\right]\,dr - \frac{1} {2}\left[\int\limits_0^{R_\infty} (\Psi_1^2(r,t)
         + \Psi_2^2(r,t))dr -1 \right]  = 0\,,
    \end{gather}
\end{subequations}
\noindent where $I$ is 2$\times$2 unit matrix.

In order to solve the PDE evolution problem \eqref{def}, \eqref{Nv} numerically we have used a standard finite-difference technique. Let us $t^0, t^1,\ldots,t^k, \ldots$, $t^0=0$, $t^{k+1}-t^k=\tau^k$ is a given discretization of the ``time'' $t$. We assume the step $0<\tau^k \le 1$ is properly specified. Then we use the Euler scheme in order to obtain the next $(k+1)$-th approximation to the exact solutions for the eigenfunction $\Psi^{k+1} \equiv \Psi(r,t^{k+1})$ and the eigenvalues $\lambda^{k+1} \equiv \lambda(t^{k+1})$:
\begin{subequations}\label{kP1}
    \begin{gather}
        \Psi^{k+1}(r)=\Psi^k(r)+\tau^k w^k(r),\\
        \lambda^{k+1}=\lambda^k+\tau^k (\mu^k-\lambda^{k}).
    \end{gather}
\end{subequations}
\noindent It is supposed that $\Psi^k(r)$ and $\lambda^k$ are
known  at the $k^{th}$ iteration step, whereas the $\Psi^0(r)$ and
$\lambda^0$ are some properly chosen initial data.

At the each step in evolution $k$ the iterative corrections
$w^k(r)$ are calculated by the following way:
\begin{eqnarray}
w^{k}=\zeta^k + \mu^k \, \eta^k,
\end{eqnarray}
where the functions $\zeta^k \equiv \left(\zeta^k_1,
\zeta^k_2\right)^T$  are solutions of the problem (for sake of
simplicity we will henceforth omit the number of iterations $k$):

\begin{subequations}\label{Nv1}
    \begin{gather}
    -\zeta\,''_1 + (U_{11}+\lambda)\, \zeta_1 + U_{12} {\zeta_2} =
    \Psi\,''_1 - U_{11} \Psi_1 - U_{12} \Psi_2,\\
    -\zeta\,''_2 + (U_{22}+\lambda)\, \zeta_2 + U_{12} {\zeta_1} =
    \Psi\,''_2 - U_{11} \Psi_2 - U_{12}\Psi_1,\\
    \zeta_1(0)=-\Psi_1(0), \quad \zeta_2'(0)=-\Psi_2'(0),\\
     \zeta\,'_1(R_{\infty}) + \sqrt{\lambda}\,\zeta_1 (R_{\infty})=
   - \Psi\, '_1(R_{\infty})-\frac{\sqrt{\lambda}\,}{2}\,\Psi_1(R_{\infty}), \\
\label{N-GLR-v1}
    \zeta\,'_2(R_{\infty}) + \sqrt{\lambda}\,\zeta_2 (R_{\infty})=
   - \Psi\, '_2(R_{\infty})-\frac{\sqrt{\lambda}\,}{2}\,\Psi_2(R_{\infty}),
   \end{gather}
\end{subequations}
whereas the functions $\eta^k \equiv \left(\eta^k_1, \eta^k_2\right)^T$ are
solutions of the problem:
\begin{subequations}\label{Nv2}
    \begin{gather}
    -\eta\,''_1 + (U_{11} + \lambda)\,\eta_1 + U_{12} \eta_2 =-\Psi_1,\\
    -\eta\,''_2 + (U_{22} + \lambda)\,\eta_2 + U_{12} \eta_1 =-\Psi_2,\\
    \eta_1(0)=0,\quad \eta_2(0)=0,\\
    \eta\,'_1(R_{\infty}) + \sqrt{\lambda}\eta_1 (R_{\infty})=
   -\frac{1}{2\sqrt{\lambda}}\Psi_1(R_{\infty}), \\
   \label{N-GLR-v2}
    \eta\,'_2(R_{\infty}) + \sqrt{\lambda}\eta_2 (R_{\infty})=
   -\frac{1}{2\sqrt{\lambda}}\Psi_2(R_{\infty}).
  \end{gather}
\end{subequations}

We used the finite difference method for the discretization in
respect to the spatial variable $r$. The corresponding discrete
linear problem \eqref{Nv1} and \eqref{Nv2}  has been solved by a
matrix sweat method. After that the value of the parameter $\mu^k$
was calculated by means of the expression \eqref{N-normaR}
\begin{eqnarray}\label{nu-k}
    \mu =\frac{1 - \int\limits_0^{R_\infty} ({\Psi_1}^2+{\Psi_2}^2)\,dr -
    2\int \limits_0^{R_\infty}(\Psi_1 \zeta_1 + \Psi_2 \zeta_2)\,dr}
     {2\,\int\limits_0^{R_\infty}( \Psi_1 {\eta}_1 + \Psi_2 {\eta}_2)\,dr}\,.
\end{eqnarray}

After the $k$~-th approximation, $\Psi^k$ and $\lambda^k$ are
found, the next, $k+1$~-th one, $\Psi^{k+1}$ and $\lambda^{k+1}$
are calculated according to the (\ref{kP1}). The iteration process
goes on until the fulfillment of the inequality $\delta^k \leq
\epsilon$, where $\epsilon
> 0$ is a given small value, and the discrepancy $\delta^k$ is defined
by the following relation:
$$ \delta^k = ||\varphi(\lambda^k, \Psi^k)|| \equiv \max_{r\in[0,R_\infty]}\,|
\varphi(\lambda^k, \Psi^k)|\,.$$
The best convergence of the method was achieved for the following choice
\cite{Pyzunin2} of the step $\tau^k$, $k>0$:
\begin{eqnarray}
    \tau^k    = \begin{cases}
     {\rm min}\left(1,\tau^{k-1}\,{\delta^{k-1}}/{\delta^k}\right),
     &\delta^k \le \delta^{k-1}, \cr \cr
    {\rm max} \left(\tau^0, \tau^{k-1}\, {\delta^{k-1}}/{\delta^k}\right),
    &\delta^k > \delta^{k-1}\,,
    \cr
    \end{cases}
\end{eqnarray}
for some given $\tau_0$ which depends on the initial approximation $z_0$.

In order to check and compare our results, we also have solved the
following non self-adjoint Sturm-Liouville problem
\begin{equation} \label{eq1}
    - \chi\,'' + P(r)\,\chi\,' + Q(r)\, \chi = -\lambda \chi\,, \quad 0 < r < \infty \,,
\end{equation}
\noindent which follows from the perturbation  functional
\eqref{Eaction}. Here $P(r)$ and $Q(r)$ are $2\times 2$ matrices,
and $P(r)$ is an antisymmetrical one. The boundary conditions are
analogous to the conditions \eqref{GrC}.

We used the collocation method for the discretization  of
appropriate linearized boundary problems at each iteration step.
The accuracy of appropriate difference scheme on the analytical
grid with the step $h$ is $O(h^2)$. The choice of the iteration
step $\tau_k$ is realized by the Ermakov-Kalitkin formula
\cite{Pyzunin2}.

\section{Numerical results}
\hspace*{\parindent} To obtain the eigenmodes of the regular
static YMD solutions the following problems have been solved:
\begin{itemize}
    \item[a)] the static background solutions, labelled by the
number $N$ of the nodes of YM function, which are solutions to the
boundary-value problem \eqref{stat}, \eqref{bcstat};
  \item[b)] the eigenmodes, which are solutions to the
self-adjoint Sturm-Liouville problem
\eqref{ShturLiuv} -- \eqref{norma} or (equivalently) to the non
self-adjoint Sturm-Liouville problem \eqref{eq1}.
\end{itemize}
\noindent Both the problems a) and b) were solved using the continuous
analogue of Newton's method.

The semi-infinity interval $[0,\infty)$ was changed  to the
interval $[0,R_\infty]$ for the numerical reasons. The influence
of the ``actual infinity'' $R_\infty$ was studied by the
establishment method. The results for the case $N = 1$ on the mesh
exponentially condensed to the origin $r = 0$, are shown in Table
 \ref{tabl}.
\begin{table}[h!]
    \caption{\label{tabl} Influence of the parameter $R_\infty$ on the eigenvalue $\lambda_1^1$}
    \begin{center}
        \begin{tabular}{|c|c|c|}
            \hline
            $R_\infty$ & $NM$ & $\lambda_1^1 \equiv -\omega^2$ \\
            \hline
            50   & 101  & $0.0905620823346630$ \\
            100  & 201  & $0.0905653977396295$ \\
            200  & 401  & $0.0905657469757358$ \\
            400  & 801  & $0.0905657826471273$ \\
            800  & 1601 & $0.0905657861599309$ \\
            1600 & 3201 & $0.0905657864933593$ \\
            \hline
        \end{tabular}
    \end{center}
\end{table}
Here $NM$ is the number of elements of the mesh. It is evident
that each doubled value of $R_\infty$ leads to the establishment
at least one significant digit after decimal point.

The one of the specific features of the problem is the rapid
increasing of $R_\infty$ as the number $N$ (which labels the
corresponding background solution) grows.
\begin{table}[h!]
    \caption{\label{tab2} First eigenvalues $\lambda_N^1$ and appropriate parameters
    $R_\infty$ and $b$.}
    \begin{center}
        \begin{tabular}{|c|c|c|c|}
            \hline
            N   & $\lambda^1_N$& $R_\infty$ & $b$  \\
            \hline
            1  &  $9.0566\times 10^{-2}$ & $2 \times 10^{3}$& $ 1.043320582$ \\
            2  &  $7.5382\times 10^{-2}$ & $2 \times 10^{5}$& $ 1.414072399$ \\
            3  &  $4.9346\times 10^{-2}$ & $2 \times 10^{7}$& $ 1.500007215$ \\
            4  &  $4.3455\times 10^{-2}$ & $1 \times 10^{8}$& $ 1.515017863$ \\
            5  &  $4.2434\times 10^{-2}$ & $1 \times 10^{9}$& $ 1.517493316$ \\
            6  &  $4.2266\times 10^{-2}$ & $3 \times 10^{10}$&$ 1.517897653$ \\
        $\infty$  &  $\approx 4.22\times 10^{-2}$ &            &$\approx 1.518$\\
            \hline
        \end{tabular}
    \end{center}
\end{table}
Table~\ref{tab2} shows sufficient values of $R_\infty$ needed for
calculation of the unstable eigenmodes for the background
solutions with $N=1,...,6$.

 In Table 2 the corresponding first eigenvalues
$\{\lambda_N^i\}_{i=1}^N$ are shown, which correspond to the
background solutions with $N=1,2,..6$.

All the values of $b$ are obtained numerically by means of the
comparison of the background solutions to the boundary-value
problem \eqref{stat}, \eqref{bcstat} and the local expansions
\eqref{expans0}. In Ref. \cite{LM} it is shown that the main
background solution parameters tend to some limit values  as $N\to
\infty$. This limit solution can be characterized by the parameter
$b_{\infty}\approx 1.518$. This allows us to determine the first
limit eigenvalue using an extrapolation of $\lambda_N^1$ as a
function of $N$: $\lambda^1_{\infty}\approx 4.22\times 10^{-2}$ (See
Table \ref{tab2}).

 In  Table~\ref{tab3} all the eigenvalues $\{\lambda_N^i\}_{i=1}^N$, which correspond to the background solutions with $N=1,2,3,4$, are presented. First three eigenvalues for the background $N=4$ solutions are also presented. For a given background solution,
labelled by the number $N$, the fast decreasing of the value
$\lambda_N^i$ as $i$ increases is obvious. Hence, only the main
(minimal) eigenmodes are presented below.

\begin{table}[h!]
    \caption{\label{tab3} Eigenvalues $\{\lambda_N^i\}_{i=1}^N$ .}
      \begin{center}
            \begin{tabular}{|l|c|c|c|c|}
                \hline
     $N$ & $\lambda_N^1$& $\lambda_N^2$& $\lambda_N^3$& $\lambda_N^4$\\ [0.1cm]
                \hline
      1       & $9.0566\times 10^{-2}$&                         &  &   \\
      2       & $7.5382\times 10^{-2}$ & $2.0742\times 10^{-4}$&  &   \\
      3       & $4.9346\times 10^{-2}$ & $1.4957\times 10^{-4}$&  $1.9622\times 10^{-7}$ &    \\
      4       & $4.3455\times 10^{-2}$ & $5.9905\times 10^{-5}$&  $1.3278\times 10^{-7}$& $\sim10^{-9}$   \\
                \hline
            \end{tabular}
    \end{center}
\end{table}

\begin{figure}[h!]
    \centering
    \includegraphics[width=11.cm]{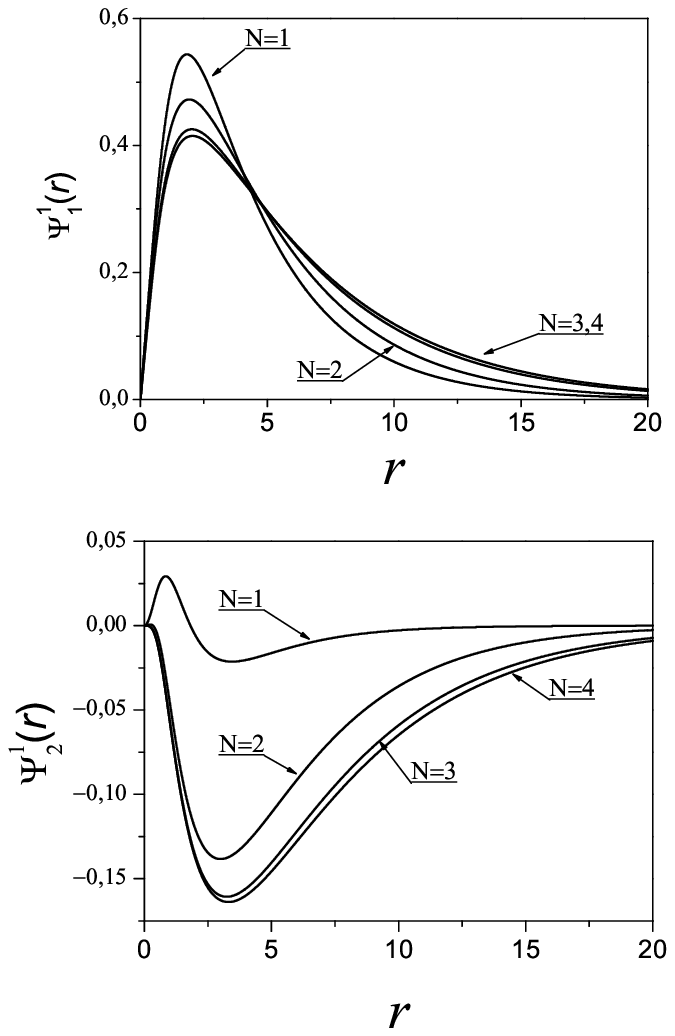}
    \caption{ \footnotesize  Eigenfunctions $\Psi_1^1$ and
 $\Psi_2^1$ for $N=1,2,3,4$.} \label{fig:3}
\end{figure}

\begin{figure}[h!]
    \centering
    \includegraphics[width=11.cm]{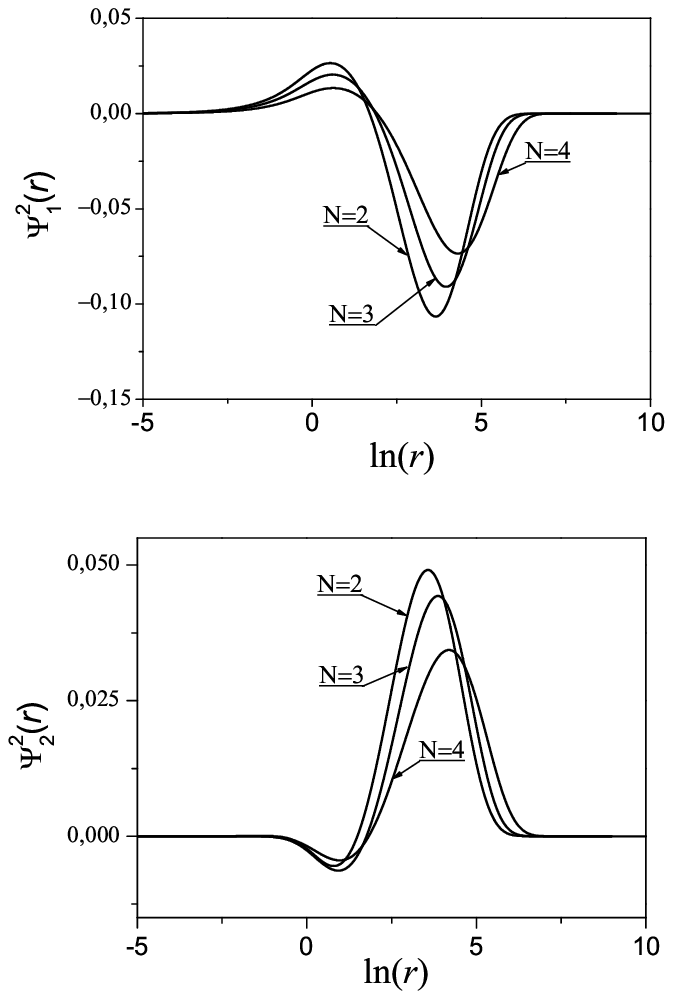}
    \caption{ \footnotesize  Eigenfunctions $\Psi_1^2$ and
 $\Psi_2^2$ for $N=2,3,4$.} \label{fig:4}
\end{figure}

\begin{figure}[h!]
    \centering
    \includegraphics[width=11.cm]{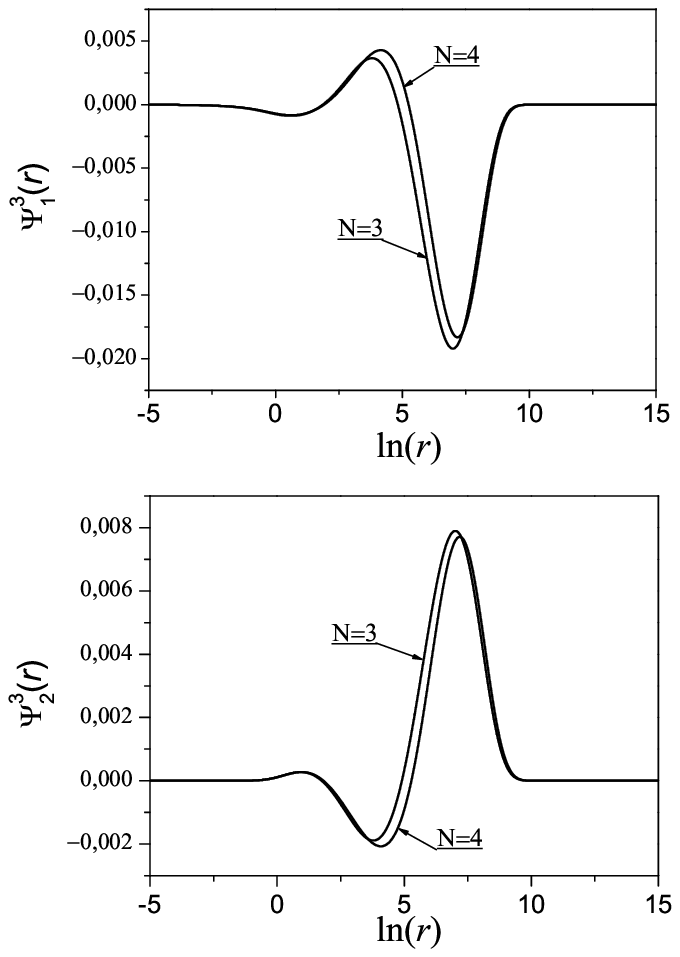}
    \caption{ \footnotesize  Eigenfunctions $\Psi_1^3$ and
 $\Psi_2^3$ for $N=3,4$.} \label{fig:5}
\end{figure}

The eigenfunctions $\Psi_1^1(r)$ and $\Psi_2^1(r)$ for $N=1,2,3,4$
are shown in Fig.\ \ref{fig:3}. The eigenfunctions $\Psi_1^2(r)$
and $\Psi_2^2(r)$ for $N=2,3,4$ are shown in Fig.\ \ref{fig:4}.
The eigenfunctions $\Psi_1^3(r)$ and $\Psi_2^3(r)$ for $N=3,4$ are
shown in Fig.\ \ref{fig:5}.

\eject

\section{Conclusions and outlook}

\hspace*{\parindent}  We have found the eigenvalues and the
eigenfunctions for the matrix Sturm-Liouville problem which are
proper unstable modes of the regular static spherically symmetric
solutions in the coupled system of Yang-Mills-dilaton equations.
The efficiency of the continuous analogue of Newton's method
(CANM), which was used to solve the corresponding numerical
problem, is demonstrated. The CANM method, being an universal tool
for numerical solution of both boundary value and Sturm-Liouville
problems, is described in detail.

 The obtained unstable modes are used as initial data in the
static solutions decay problem in a nonlinear regime. This task is
in progress now and will be reported soon \cite{decay}. The
mentioned nonlinear decay problem arises in a wide class of
mathematical physics problems, inspired by the recent progress in
the singularity formation understanding in nonlinear wave
equations and in massless fields collapse problem.

\subsection*{Acknowledgments}

 Discussions with E. P. Zhidkov and I. V.
Puzynin are greatly acknowledged. We are grateful to J. Bu\v{s}a
and I. Pokorn\'y for technical support and helpful discussions.
The work is supported in part by the RFBR Grant N 02-01-00606.

\appendix
\section{Appendix}

To facilitate the readers, the continuous analogue of Newton's
method (CANM) is presented briefly below.

Let us consider the nonlinear equation
\begin{equation}\label{eq:a1}
    \chi(x) = 0,
\end{equation}
where the nonlinear operator $y = \chi(x)$ maps the Banach space
$X$ (the element $x \in X$) into the Banach space $Y (y \in Y)$.
Solution to the eq.\ \eqref{eq:a1} means finding such elements
$x^* \in X$, which are mapped by the operator $\chi$ into the
trivial element of space $Y$.

If we suppose that $x_k$ is known approximation to the sought
solution at the $k$th iteration stage, then the  increment $\Delta
x_k$ is computed by the formula $$\Delta x_k = \psi(x_k), \>
x_{k+1} = x_k + \Delta x_k,\> k =0,1,2,...\>, \hbox{where}\>\>
x_0\> \hbox{- given element}.$$

The used method of iteration determines the manner of constructing
the function $\psi(x)$. For example, in the case of Newton's
method $\psi(x) = -[\chi^\prime (x)]^{-1} \chi(x)$ where
$\chi^\prime (x)$ is a linear operator, it is the  Frech{\'e}t
derivative of function $\chi(x)$. For each iteration process of
the kind mentioned above one can build a continuous analogue
introducing a continuous parameter $t,$ $0 \le t < \infty$ instead
the discrete variable $k$, $(k = 0,1,2,...).$ Further we suppose
the smooth dependence $x = x(t)$ and introduce the derivative
$\frac{d}{d t} x(t)$ instead of the increment $\Delta x_k$. In
this way we obtain the differential equation
\begin{equation}
    \frac{d}{dt} x(t) = - \psi[(x(t)], \quad x(0)= x_0. \label{eq:a2}
\end{equation}
Thus, the solving the original eq.(\ref{eq:a1}) is realized by the
solving of the above Cauchy problem (\ref{eq:a2}) on the positive
half-axis $0 \le t < \infty$.

There have been proved numerous theorems (see \cite{gavurin},
\cite{Pyzunin0, Pyzunin2}) related to the convergence to the
isolated solution $x^*$ of the continuous analogues of various
iteration methods.

As a particular case of the continuous analogue of Newton's method
(CANM) we can present the eq.({\ref{eq:a2}) as
\begin{equation}
    \frac{d}{d t} \chi [x(t)] = - \chi [ x(t)], \quad x(0) = x_0. \label{eq:a3}
\end{equation}
From here we obtain the first integral
$$\chi[x(t)] = {\rm e}^{-t} \chi (x_0).$$

If the function $\chi(x)$ is smooth and the operator $[\chi^\prime
(x)]^{-1}$ is bounded in the vicinity of initial approximation
$x_0$, then in the same vicinity there exists an isolated root
$x^*$ of the eq.(\ref{eq:a1}) and $\lim_{t \to \infty} x(t) \to
x^*$.

For example, let us consider the convergence conditions of CANM
for the following simple non-linear boundary value problem:
\begin{equation}
    \chi(y) \equiv \{y^{\prime\prime} + f(x,y),\> y(0),\> y(1)\} = 0,
    \quad x \in (0,1) \label{eq:a4}\\.
\end{equation}

\begin{theorem}\label{jmp}
    Let the solution of BVP (\ref{eq:a4}) exist and can be localized. Furthermore
    \begin{description}
        \item[(i)] the function $f(x,y)$ is smooth in some domain $D$;
        \item[(ii)] the boundary value problem
        $$v^{\prime\prime} + f^\prime_y (x,y) v = 0,\quad v(0) = v(1) = 0$$
has only a trivial solution for each smooth function $y(x) \in D$;
        \item[(iii)] $\| y_0^{\prime\prime} + f(x,y_0) \| \le \epsilon$ where
        $\epsilon > 0$ is little enough, and function $y_0 (x) $, smooth in domain $D$
         is an initial approximation of the sought solution $y^* (x)$.
    \end{description}

Then the system with respect to functions $y(x,t)$ and $v(x,t)$
    \begin{equation*}
        v_{xx}^{\prime\prime} + f_y^\prime (x,y) v = -\left[y_{xx}^{\prime\prime} +
        f(x,y)\right], \quad y_t^\prime = v \label{eq:a5}
    \end{equation*}
with boundary conditions
    \begin{equation*}
        v(0,t) = v(1,t) =0, \label{eq:a6}
    \end{equation*}
    and initial condition
    $$y(x,0) = y_0 (x)$$
has on half-strip $s = \{(x,t)\!: 0 \le x \le 1, 0 \le t < \infty
\}$ a unique solution, subjected to the condition
    $$ \lim_{t \to \infty} \|y(x,t) - y^* (x) \|_{C^2[0,1]} = 0.$$
\end{theorem}

The most simple method for approximated integration of the problem
(\ref{eq:a2}) is the Euler method. Let us build the set $t_k,\> k
= 0,1,2,...$ and $\tau_k = t_{k+1} - t_k$. Then the following
sequence of linear problems is reached:
$$
\chi^\prime (x_k) v_k = - \chi (x_k),\quad x_{k+1} = x_k + \tau_k
v_k, \quad k = 0,1,2,..., $$ where $x_0$ is a predetermined
element. When the parameter $\tau_k \equiv 1$, we obtain the
classical Newton's method.

\end{document}